\documentclass{PoS}

\newcommand{\Pom}{\mathbb{P}}
\newcommand{\Ode}{\mathbb{O}}
\newcommand{\Reg}{\mathbb{R}}

\renewcommand\slash[1]{\not \! #1}

\newcommand{\bp}{\mbox{\boldmath $p$}}
\newcommand{\bk}{\mbox{\boldmath $k$}}

\title{Searching for odderon exchange in exclusive $pp \to pp \phi \phi$ reaction}

\ShortTitle{Searching for odderon exchange in exclusive $pp \to pp \phi \phi$ reaction}

\author{Piotr Lebiedowicz\\
        Institute of Nuclear Physics Polish Academy of Sciences, 
        Radzikowskiego 152, PL-31342 Cracow, Poland\\
        E-mail: \email{Piotr.Lebiedowicz@ifj.edu.pl}}

\author{Otto Nachtmann\\
        Institut f{\"u}r Theoretische Physik, Universit{\"a}t Heidelberg,
        Philosophenweg 16, D-69120 Heidelberg, Germany\\
        E-mail: \email{O.Nachtmann@thphys.uni-heidelberg.de}}

\author{\speaker{Antoni Szczurek}\thanks{This work was partially supported by
the Polish National Science Centre Grant No. 2018/31/B/ST2/03537.}\\
        Institute of Nuclear Physics Polish Academy of Sciences,
        Radzikowskiego 152, PL-31342 Cracow, Poland\\
        E-mail: \email{Antoni.Szczurek@ifj.edu.pl}}

\abstract{We discuss the possibility to use the $pp \to pp \phi \phi$ process 
in identifying the odderon exchange.
So far there is no unambiguous experimental evidence for the odderon,
the charge conjugation $C = -1$ counterpart of the $C= +1$ pomeron.
Last year results of the TOTEM collaboration suggest 
that the odderon exchange can be responsible for a disagreement 
of theoretical calculations and the TOTEM data for 
elastic proton-proton scattering. 
Here we present recent studies for central exclusive production (CEP) of $\phi \phi$ pairs 
in proton-proton collisions.
We consider the pomeron-pomeron fusion to $\phi \phi$ ($\Pom \Pom \to \phi \phi$)
through the continuum processes, due to the $\hat{t}$- and $\hat{u}$-channel 
reggeized $\phi$-meson, photon, and odderon exchanges,
as well as through the $s$-channel resonance process ($\Pom \Pom \to f_{2}(2340) \to \phi \phi$).
This $f_{2}$ state is a candidate for a tensor glueball.
The amplitudes for the processes are formulated
within the tensor-pomeron and vector-odderon approach.
Some model parameters are determined from the comparison 
to the WA102 experimental data.
The odderon exchange is not excluded by the WA102 data 
for high $\phi \phi$ invariant masses.
The measurement of large $M_{\phi \phi}$ or $\rm{Y_{diff}}$ events
at the LHC would therefore suggest presence of the odderon exchange.}

\FullConference{
European Physical Society Conference on High Energy Physics - EPS-HEP2019 -\\
			10-17 July, 2019\\
			Ghent, Belgium}

\begin{document}

\section{Introduction}

Diffractive studies are one of the important parts of the physics program
for the RHIC \cite{Sikora:2018cyk} and LHC experiments \cite{CMS, Staszewski:2011bg}. 
A particularly interesting class is the central-exclusive-production (CEP) processes, 
where all centrally produced particles are detected. 
In recent years, there has been a renewed interest in exclusive production 
of $\pi^+ \pi^-$ pairs at high energies
related to successful experiments by 
the CDF \cite{Aaltonen:2015uva} and the CMS \cite{CMS} collaborations.
Preliminary results of similar CEP studies have been presented by the ALICE
and LHCb collaborations at the LHC.
These measurements are important in the context of resonance production, 
in particular, in searches for glueballs.
For a feasibility studies for the $p p \to p p \pi^+ \pi^-$ process with 
tagging of the scattered protons as carried out 
for the ATLAS and ALFA detectors see \cite{Staszewski:2011bg}.

In \cite{Ewerz:2013kda} the tensor-pomeron and vector-odderon concept 
was introduced for soft reactions.
In this approach, the $C = +1$ pomeron and the reggeons 
$\Reg_{+} = f_{2 \Reg}, a_{2 \Reg}$ are treated as effective
rank-2 symmetric tensor exchanges
while the $C = -1$ odderon and the reggeons 
$\Reg_{-} = \omega_{\Reg}, \rho_{\Reg}$ are treated 
as effective vector exchanges.
For these effective exchanges a number of propagators and vertices, 
respecting the standard rules of quantum field theory, 
were derived from comparisons with experiments.
This allows for an easy construction of amplitudes for specific processes.
Applications of the tensor-pomeron and vector-odderon ansatz
were given for photoproduction of pion pairs in \cite{Bolz:2014mya}
and for a number of central-exclusive-production (CEP) reactions in $pp$ collisions 
in \cite{Lebiedowicz:2013ika,Lebiedowicz:2014bea,Lebiedowicz:2016ioh,
Lebiedowicz:2016zka,Lebiedowicz:2016ryp,Lebiedowicz:2018sdt,
Lebiedowicz:2018eui,Lebiedowicz:2019por,Lebiedowicz:2019jru}.
In \cite{Ewerz:2016onn} the helicity structure of small-$|t|$ proton-proton elastic scattering
was considered in three models for the pomeron: tensor, vector, and scalar.
Only the tensor ansatz for the pomeron was found to be compatible with 
the high-energy experiment on polarized $pp$ elastic scattering \cite{Adamczyk:2012kn}.

So far there is no unambiguous experimental evidence for the odderon ($\Ode$),
the charge conjugation $C = -1$ counterpart of the $C= +1$ pomeron,
introduced on theoretical grounds in \cite{Lukaszuk:1973nt}.
A hint of the odderon was seen in ISR results \cite{Breakstone:1985pe}
as a small difference between the differential cross sections
of elastic proton-proton ($pp$) and proton-antiproton ($p \bar{p}$) scattering
in the diffractive dip region at $\sqrt{s} = 53$~GeV.
Recently the TOTEM Collaboration has published data from high-energy elastic
$pp$ scattering experiments at the LHC \cite{Antchev:2017yns}.

\begin{figure}
\begin{center}
(a)\includegraphics[width=0.3\textwidth]{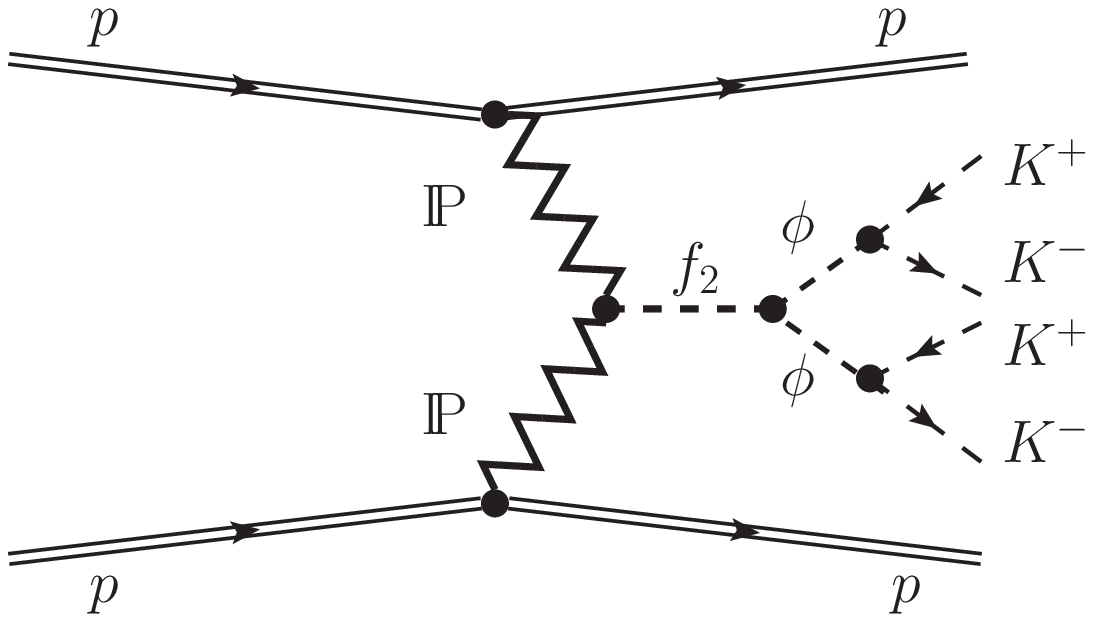} 
(b)\includegraphics[width=0.3\textwidth]{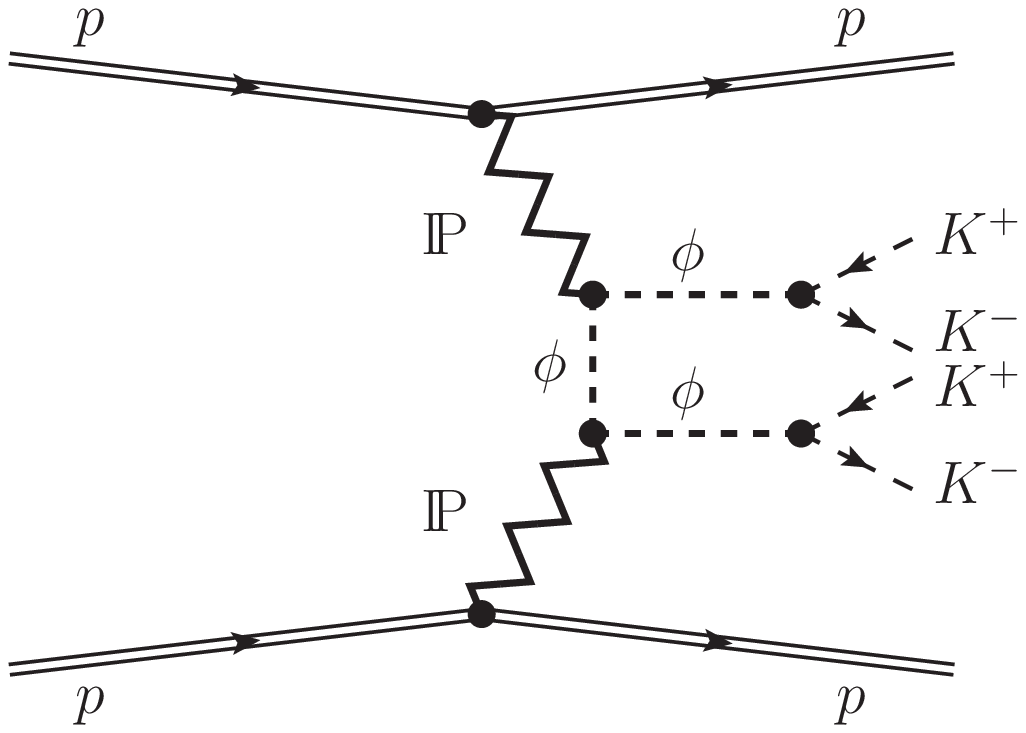}   
(c)\includegraphics[width=0.3\textwidth]{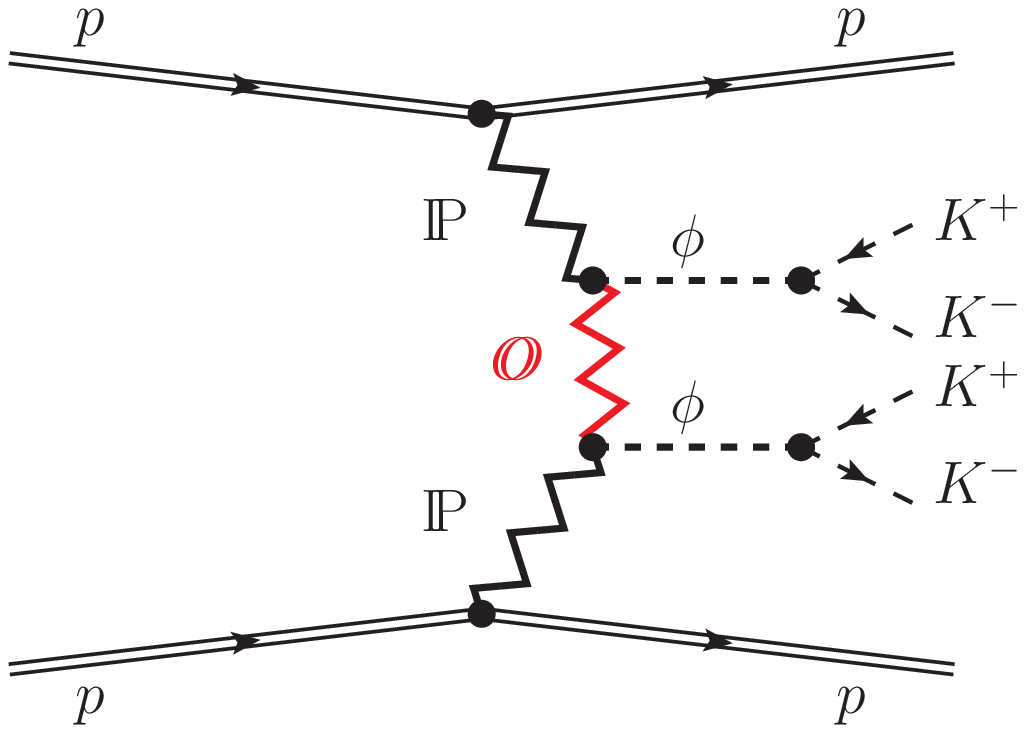}
\caption{The Born-level diagrams for double pomeron central exclusive $\phi \phi$ production 
and their decays into $K^+ K^- K^+ K^-$:
(a) $\phi\phi$ production via an $f_{2}$ resonance.
Other resonances, e.g. of $f_{0}$- and $\eta$-type, can also contribute here.
(b) and (c) continuum $\phi\phi$ production via an intermediate
$\phi$ and odderon ($\Ode$) exchanges, respectively.
$\Pom$-$\gamma$-$\Pom$ and $\Ode$-$\Pom$-$\Ode$ contributions are also possible but negligibly small.}
\label{fig:4K_diagrams}
\end{center}
\end{figure}

As was discussed in \cite{Schafer:1991na} 
exclusive diffractive $J/\psi$ and $\phi$ production from the pomeron-odderon fusion
in high-energy $pp$ and $p\bar{p}$ collisions 
is a direct probe for a possible odderon exchange.
For a nice review of odderon physics see \cite{Ewerz}.
In the diffractive production of $\phi$ meson pairs 
it is possible to have pomeron-pomeron fusion 
with intermediate $\hat{t}/\hat{u}$-channel odderon exchange \cite{Lebiedowicz:2019jru};
see the corresponding diagram in Fig.~\ref{fig:4K_diagrams}~(c).
Thus, the $pp \to pp \phi \phi$ reaction
is a good candidate for the $\Ode$-exchange searches, 
as it does not involve the coupling of the odderon to the proton \cite{Lebiedowicz:2018sdt}.

Studies of different decay channels in central exclusive production 
would be very valuable also in the context of identification of glueballs.
One of the promising reactions is $p p \to p p \phi \phi$
with both $\phi \equiv \phi(1020)$ mesons decaying into the $K^{+} K^{-}$ channel.
Structures in the $\phi \phi$ invariant-mass spectrum were observed by several experiments
\cite{Etkin:1987rj, Aston:1990wf, Barberis:1998bq}.
Three tensor states, $f_{2}(2010)$, $f_{2}(2300)$, and $f_{2}(2340)$, 
observed previously in \cite{Etkin:1987rj},
were also observed in the radiative decay $J/\psi \to \gamma \phi \phi$ \cite{Ablikim:2016hlu}.
The nature of these resonances is not understood at present.
According to lattice-QCD simulations, the lightest tensor glueball 
has a mass between 2.2 and 2.4~GeV, see, e.g. \cite{Morningstar:1999rf}.
The $f_{2}(2300)$ and $f_{2}(2340)$ states are good candidates to be tensor glueballs.

\section{A sketch of formalism}

In \cite{Lebiedowicz:2019jru} we considered the CEP of four charged kaons
via the intermediate $\phi \phi$ state.
Explicit expressions for the $pp \to pp \phi \phi$ amplitudes 
involving the pomeron-pomeron fusion to $\phi \phi$ ($\Pom \Pom \to \phi \phi$)
through the continuum processes, due to the $\hat{t}$- and $\hat{u}$-channel 
reggeized $\phi$-meson, photon, and odderon exchanges,
as well as through the $s$-channel resonance reaction ($\Pom \Pom \to f_{2}(2340) \to \phi \phi$)
were given there.
The ``Born-level'' amplitude for the $pp \to pp \phi \phi$ reaction can be written as
\begin{eqnarray}
{\cal {M}}^{\mathrm{Born}} &=&
{\cal {M}}^{(f_{2}{\rm-exchange})} + 
{\cal {M}}^{(\phi{\rm-exchange})} + 
{\cal {M}}^{(\Ode{\rm-exchange})}\,.
\label{amp_Born}
\end{eqnarray}
Here we discuss shortly the continuum process
with the odderon exchange [see Fig.~\ref{fig:4K_diagrams}~(c)]
for the $pp \to pp \phi \phi$ reaction.
The amplitude is a sum of $\hat{t}$- and $\hat{u}$-channel amplitudes.
The $\hat{t}$-channel term (Born level) can be written as 
%
\begin{eqnarray}
{\cal M}^{(\hat{t})} 
= 
&&(-i) \bar{u}(p_{1}, \lambda_{1}) 
i\Gamma^{(\Pom pp)}_{\mu_{1} \nu_{1}}(p_{1},p_{a}) 
u(p_{a}, \lambda_{a})
i\Delta^{(\Pom)\, \mu_{1} \nu_{1}, \alpha_{1} \beta_{1}}(s_{13},t_{1})  \nonumber \\
&&\times 
i\Gamma^{(\Pom \Ode \phi)}_{\rho_{1} \rho_{3} \alpha_{1} \beta_{1}}(\hat{p}_{t},-p_{3})
\left(\epsilon^{(\phi)\,\rho_{3}}(\lambda_{3})\right)^*
i\Delta^{(\Ode)\,\rho_{1} \rho_{2}}(s_{34}, \hat{p}_{t}) \nonumber \\
&&\times 
i\Gamma^{(\Pom \Ode \phi)}_{\rho_{4} \rho_{2} \alpha_{2} \beta_{2}}(p_{4},\hat{p}_{t})
\left(\epsilon^{(\phi)\,\rho_{4}}(\lambda_{4})\right)^*
i\Delta^{(\Pom)\, \alpha_{2} \beta_{2}, \mu_{2} \nu_{2}}(s_{24},t_{2}) \nonumber \\
&&\times 
\bar{u}(p_{2}, \lambda_{2}) 
i\Gamma^{(\Pom pp)}_{\mu_{2} \nu_{2}}(p_{2},p_{b}) 
u(p_{b}, \lambda_{b})\,,
\label{amplitude_t}
\end{eqnarray}
where 
$p_{a,b}$, $p_{1,2}$ and $\lambda_{a,b}, \lambda_{1,2} = \pm \frac{1}{2}$ 
denote the four-momenta and helicities of the protons and
$p_{3,4}$ and $\lambda_{3,4} = 0, \pm 1$ 
denote the four-momenta and helicities of the $\phi$ mesons, respectively.
$\hat{p}_{t} = p_{a} - p_{1} - p_{3}$,
$\hat{p}_{u} = p_{4} - p_{a} + p_{1}$, $s_{ij} = (p_{i} + p_{j})^{2}$,
$t_1 = (p_{1} - p_{a})^{2}$,
$t_2 = (p_{2} - p_{b})^{2}$.
$\Gamma^{(\Pom pp)}$ and $\Delta^{(\Pom)}$ 
denote the proton vertex function and the effective propagator,
respectively, for tensorial pomeron.
The corresponding expressions are as follows \cite{Ewerz:2013kda}:
\begin{eqnarray}
&&i\Gamma_{\mu \nu}^{(\Pom pp)}(p',p)
=-i 3 \beta_{\Pom NN} F_{1}(t) 
\left\lbrace 
\frac{1}{2} 
\left[ \gamma_{\mu}(p'+p)_{\nu} 
     + \gamma_{\nu}(p'+p)_{\mu} \right]
- \frac{1}{4} g_{\mu \nu} ( \slash{p}' + \slash{p} )
\right\rbrace\,,
\label{A4} \\
&&i \Delta^{(\Pom)}_{\mu \nu, \kappa \lambda}(s,t) =
\frac{1}{4s} \left( g_{\mu \kappa} g_{\nu \lambda} 
                  + g_{\mu \lambda} g_{\nu \kappa}
                  - \frac{1}{2} g_{\mu \nu} g_{\kappa \lambda} \right)
(-i s \alpha'_{\Pom})^{\alpha_{\Pom}(t)-1}\,,
\label{A1}
\end{eqnarray}
where $\beta_{\Pom NN} =1.87$~GeV$^{-1}$. The pomeron trajectory $\alpha_{\Pom}(t)$
is assumed to be of standard linear form (see, e.g., \cite{DDLN}):
$\alpha_{\Pom}(t) = \alpha_{\Pom}(0)+\alpha'_{\Pom}\,t$,
$\alpha_{\Pom}(0) = 1.0808$,
$\alpha'_{\Pom} = 0.25 \; {\rm GeV}^{-2}$.

Our ansatz for the effective propagator of the $C = -1$ odderon is \cite{Ewerz:2013kda}
\begin{eqnarray}
i \Delta^{(\Ode)}_{\mu \nu}(s,t) = 
-i g_{\mu \nu} \frac{\eta_{\Ode}}{M_{0}^{2}} (-i s \alpha'_{\Ode})^{\alpha_{\Ode}(t)-1}  \quad
{\rm with}
\quad M_{0} = 1~{\rm GeV}\,,\,\, \eta_{\Ode} =\pm 1\,.
\label{A12} 
\end{eqnarray}
%
Here $\alpha_{\Ode}(t) = \alpha_{\Ode}(0)+\alpha'_{\Ode}\,t$
and we choose, as an example, 
$\alpha'_{\Ode} = 0.25$~GeV$^{-2}$, $\alpha_{\Ode}(0) = 1.05$.
%
%

For the $\Pom \Ode \phi$ vertex we use an ansatz
with two rank-four tensor functions $\Gamma^{(0,2)}$ \cite{Lebiedowicz:2019jru}:
%
\begin{eqnarray}
&&i\Gamma^{(\Pom \Ode \phi)}_{\mu \nu \kappa \lambda}(k',k) =
i F^{(\Pom \Ode \phi)}((k+k')^{2},k'^{2},k^{2}) 
\left[ 2\,a_{\Pom \Ode \phi}\, \Gamma^{(0)}_{\mu \nu \kappa \lambda}(k',k)
- b_{\Pom \Ode \phi}\,\Gamma^{(2)}_{\mu \nu \kappa \lambda}(k',k) \right],\quad
\label{A15}\\
%
%
&&F^{(\Pom \Ode \phi)} ( (k+k')^{2},k'^{2},k^{2} ) = 
F((k+k')^{2})\, F(k'^{2})\, F^{(\Pom \Ode \phi)}(k^{2})\,,
\label{Fpomodephi}
\end{eqnarray}
where we take
$F(k^{2})= (1-k^{2}/\Lambda^{2})^{-1}$
%
and 
$F^{(\Pom \Ode \phi)}(m_{\phi}^{2}) = 1$.
The coupling parameters $a_{\Pom \Ode \phi}$, $b_{\Pom \Ode \phi}$
and the cutoff parameter $\Lambda^{2}$
could be adjusted
to the WA102 experimental data \cite{Barberis:1998bq}.

At low $\sqrt{s_{34}} = {\rm M}_{\phi\phi}$ the Regge type of interaction is not realistic
and should be switched off. 
To achieve this 
we multiplied the $\Ode$-exchange amplitude by a purely phenomenological factor:
%
$F_{{\rm thr}}(s_{34}) = 1 - \exp[(s_{\rm thr}-s_{34})/s_{\rm thr})]$ 
with $s_{\rm thr} = 4 m_{\phi}^{2}$.
%

The amplitude for the process shown in Fig.~\ref{fig:4K_diagrams}~(b)
has the same form as the amplitude with the $\Ode$~exchange
but we have to make the following replacements:
\begin{eqnarray}
&&i\Gamma^{(\Pom \Ode \phi)}_{\mu \nu \kappa \lambda}(k',k)
\to i\Gamma^{(\Pom \phi\phi)}_{\mu \nu \kappa \lambda}(k',k)\,,\\
&&i\Delta^{(\Ode)}_{\mu \nu}(s_{34},\hat{p}^{2}) 
\to i\Delta^{(\phi)}_{\mu \nu}(\hat{p})\,.
\end{eqnarray}
We have fixed the coupling parameters of the tensor pomeron to the $\phi$ meson 
based on the HERA experimental data for the $\gamma p \to \phi p$ reaction; see \cite{Lebiedowicz:2018eui}.
We should take into account the reggeization of the intermediate $\phi$ meson; see \cite{Lebiedowicz:2019jru}.
We take
$\alpha_{\phi}(\hat{p}^{2}) = \alpha_{\phi}(0)+\alpha'_{\phi}\,\hat{p}^{2}$,
$\alpha_{\phi}(0) = 0.1$ \cite{Collins}, and
$\alpha'_{\phi} = 0.9 \; {\rm GeV}^{-2}$.
%
%
%

In order to give realistic predictions we shall include absorption effects
calculated at the amplitude level and related to the $pp$ nonperturbative interactions.
The full amplitude includes the $pp$-rescattering corrections (absorption effects)
\begin{eqnarray}
&&{\cal {M}}_{pp \to pp \phi \phi} =
{\cal {M}}^{\mathrm{Born}} + 
{\cal {M}}^{\mathrm{absorption}}\,,\\
%
&&{\cal M}^{\mathrm{absorption}}(s,\bp_{1t},\bp_{2t})= 
\frac{i}{8 \pi^{2} s} \int d^{2} \bk_{t} \,
{\cal M}^{\mathrm{Born}}(s,\tilde{\bp}_{1t},\tilde{\bp}_{2t})\,
{\cal M}_{\mathrm{el}}^{(\Pom)}(s,-{\bk}_{t}^{2})\,,
\label{abs_correction}
\end{eqnarray}
where $\tilde{\bp}_{1t} = {\bp}_{1t} - {\bk}_{t}$ and
$\tilde{\bp}_{2t} = {\bp}_{2t} + {\bk}_{t}$.
${\cal M}_{\mathrm{el}}^{(\Pom)}$ 
is the elastic $pp$-scattering amplitude
with the momentum transfer $t=-{\bk}_{t}^{2}$.

\section{Results}

\begin{figure}
\begin{center}
\includegraphics[width=0.32\textwidth]{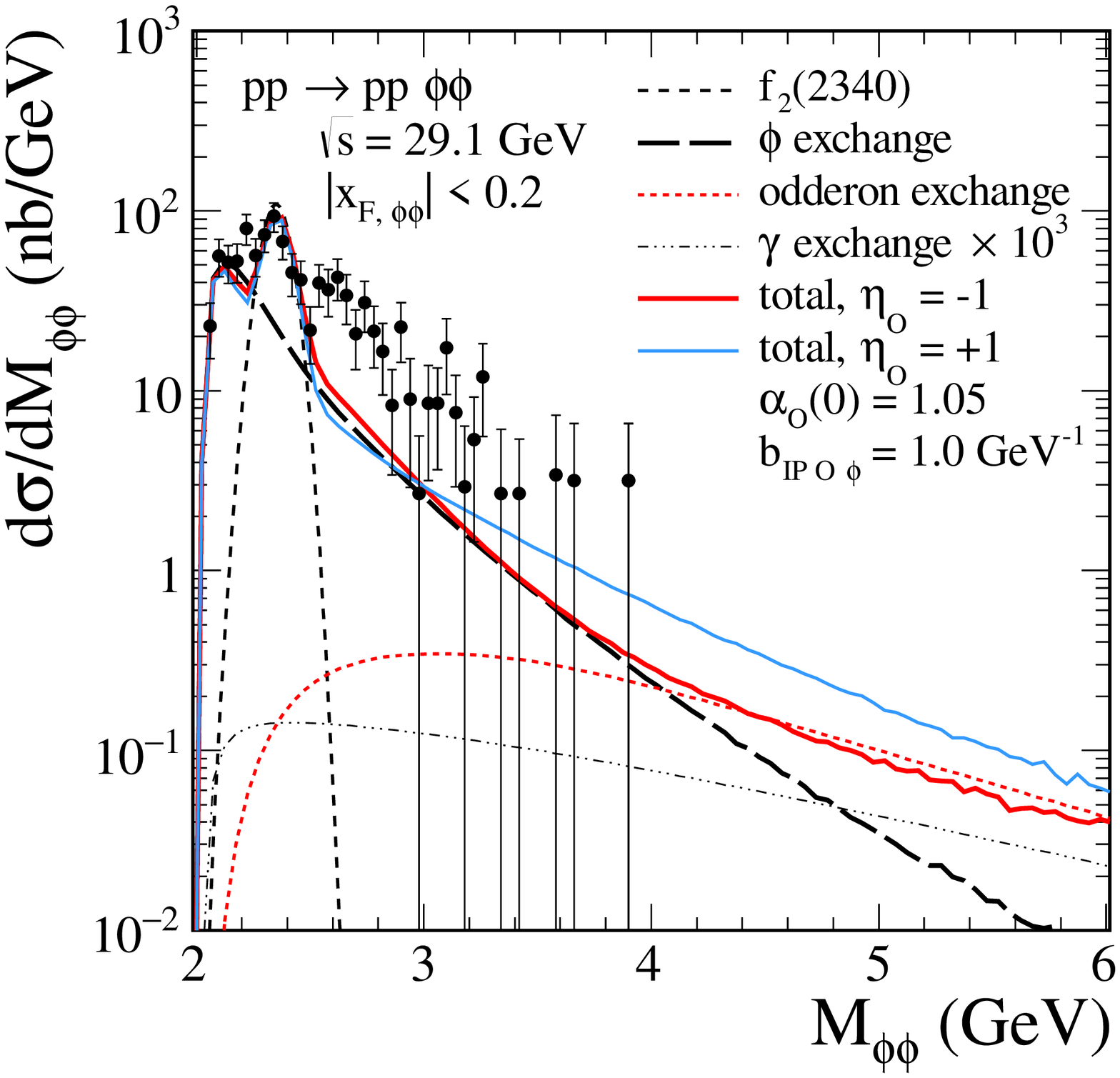}
\includegraphics[width=0.32\textwidth]{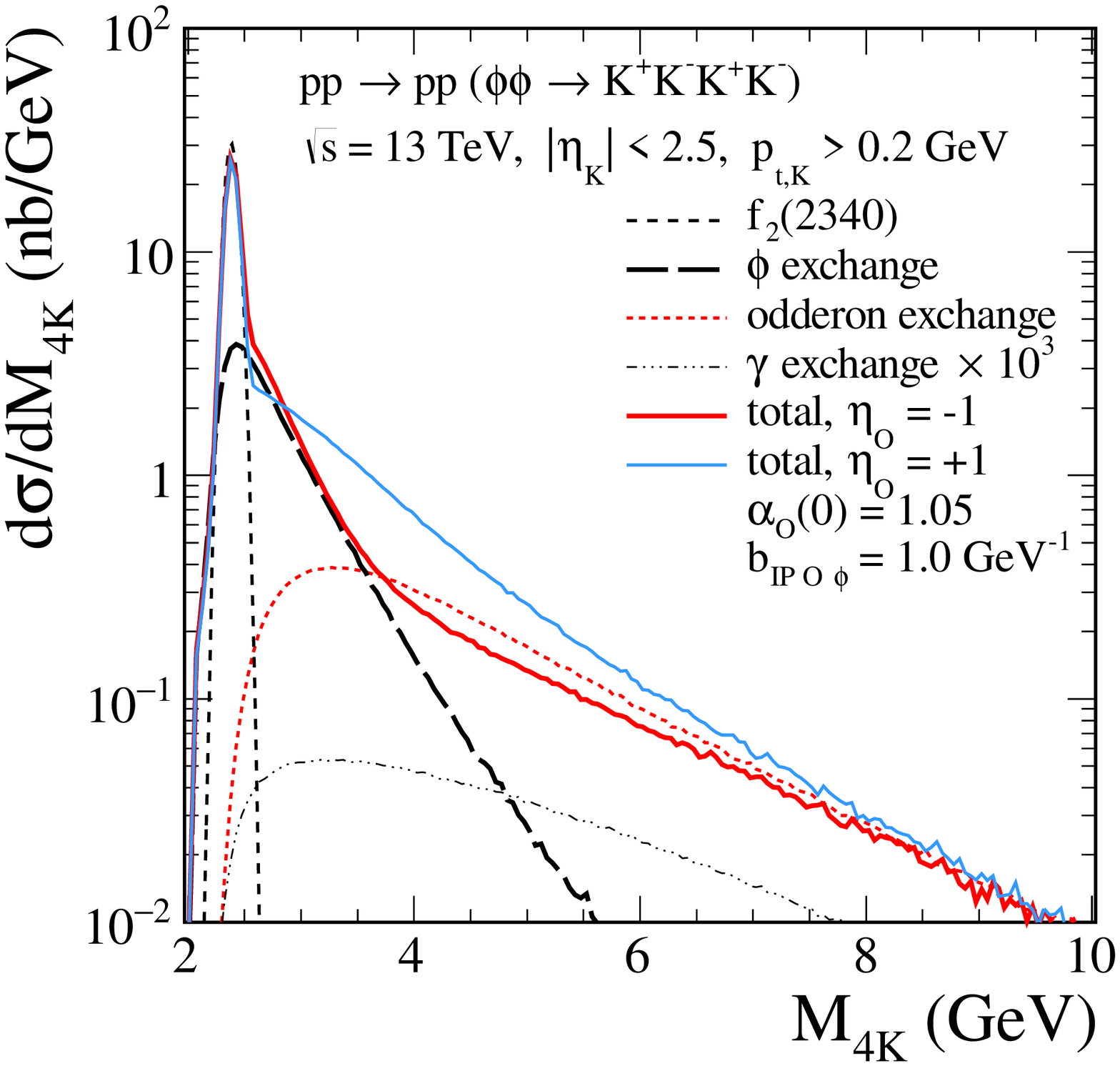}
\includegraphics[width=0.32\textwidth]{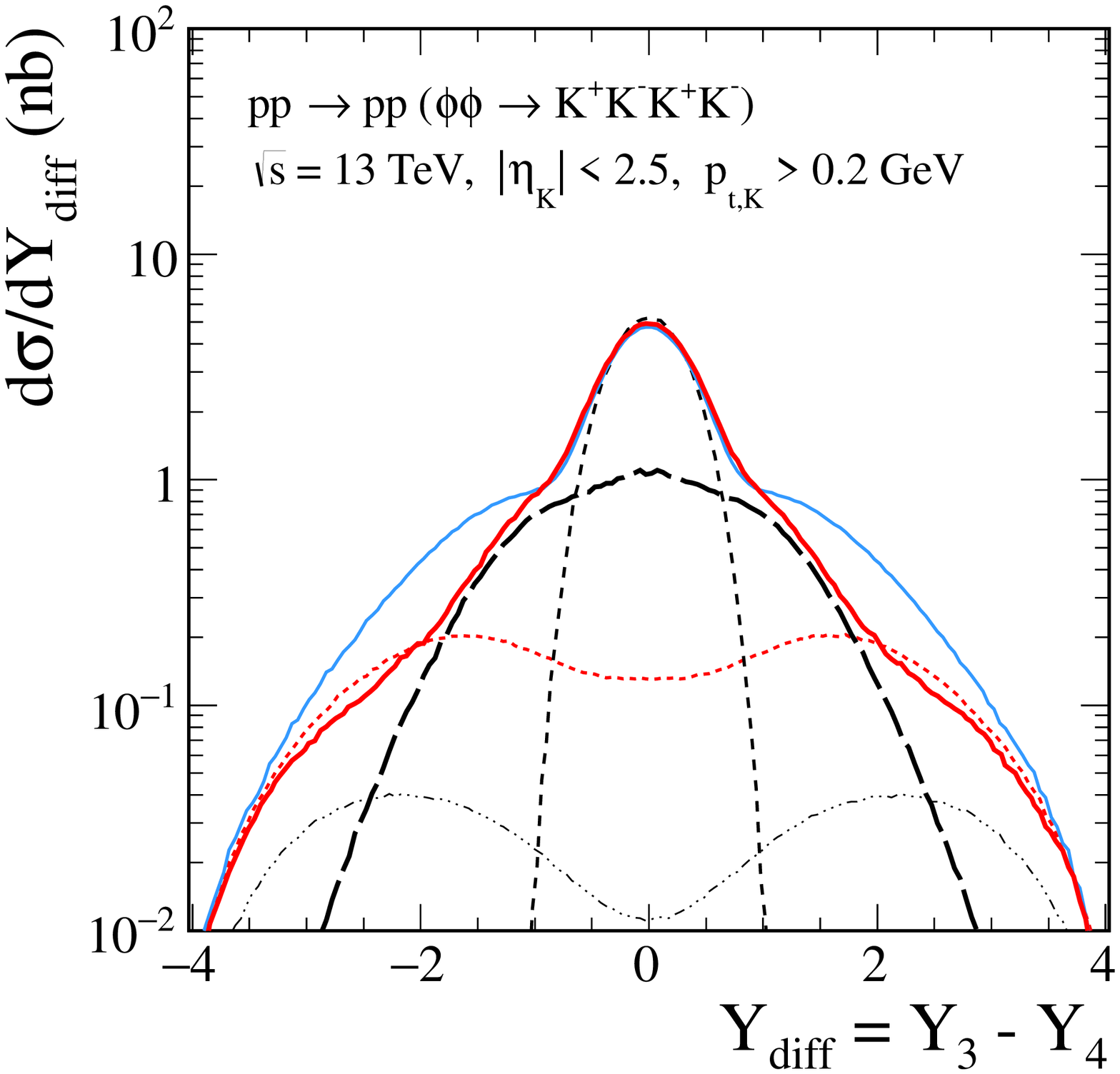}
\caption{The distributions in $\phi\phi$ invariant mass (left),
in ${\rm M}_{4K}$ (center), and in $\rm{Y_{diff}}$ (right).
The WA102 experimental data from \cite{Barberis:1998bq} are shown.
The calculations were done for $\sqrt{s} = 29.1$~GeV and $|x_{F,\phi \phi}| \leqslant 0.2$,
and for $\sqrt{s} = 13$~TeV and $|\eta_{K}| < 2.5$, $p_{t, K} > 0.2$~GeV.
The black long-dashed line corresponds to the $\phi$-exchange contribution 
and the black dashed line corresponds to the $f_{2}(2340)$ contribution.
The red dotted line represents the odderon-exchange contribution
for $a_{\Pom \Ode \phi}=0$ and $b_{\Pom \Ode \phi}= 1.0$~GeV$^{-1}$ in (\ref{A15}).
The coherent sum of all terms is shown by the red and blue solid lines
for $\eta_{\Ode} = -1$ and $\eta_{\Ode} = +1$, respectively.
The absorption effects are included in the calculations.}
\label{fig:odderon_LHC}
\end{center}
\end{figure}

Figure~\ref{fig:odderon_LHC} shows complete results
including the $f_{2}(2340)$-resonance contribution
and the continuum processes due to reggeized-$\phi$, odderon, and photon exchanges.
It is very difficult to describe the WA102 data for the $p p \to p p \phi \phi$
reaction including resonances and the $\phi$-exchange mechanism only \cite{Lebiedowicz:2019jru}.
Inclusion of the odderon exchange improves the description of the WA102 data \cite{Barberis:1998bq},
see the left panel of Fig.~\ref{fig:odderon_LHC}.
Having fixed the parameters of our quasi fit to the WA102 data
we wish to show our predictions for the LHC.
We show the results for the ATLAS (CMS) experimental conditions 
($|\eta_{K}| < 2.5$, $p_{t, K} > 0.2$~GeV).
The distribution in four-kaon invariant mass is shown in the center panel
and the difference in rapidity between the two $\phi$ mesons ($\rm{Y_{diff}}$) in the right panel.
The small intercept of the $\phi$-reggeon exchange, 
$\alpha_{\phi}(0) = 0.1$
makes the $\phi$-exchange contribution steeply falling 
with increasing ${\rm M}_{4K}$ and $|\rm{Y_{diff}}|$.
Therefore, an odderon with an intercept $\alpha_{\Ode}(0)$ around 1.0
should be clearly visible in the region of large ${\rm M}_{4K}$ and 
for large rapidity distance between the $\phi$ mesons.

\section{Conclusions}

Taking into account typical kinematic cuts for LHC experiments
in the $pp \to pp \phi \phi \to pp K^{+}K^{-}K^{+}K^{-}$ reaction
we have found that the odderon exchange contribution
should be distinguishable from other contributions
for large rapidity distance between the outgoing $\phi$ mesons 
and in the region of large four-kaon invariant masses.
Our results can be summarized in the following way:
\begin{itemize}
	\item 
CEP is particularly interesting class of processes which provides insight to unexplored soft QCD phenomena.
The fully differential studies of exclusive $pp \to pp \phi \phi$ reaction 
within the tensor-pomeron and vector-odderon approach was executed; 
for more details see \cite{Lebiedowicz:2019jru}.
Integrated cross sections of order of a few nb are obtained
including the experimental cuts relevant for the LHC experiments.
The distribution in rapidity difference of both $\phi$-mesons
could shed light on the $f_{2}(2340) \to \phi \phi$ coupling \cite{Lebiedowicz:2016ioh}, not known at present.
	\item
We find from our model that the odderon-exchange contribution
should be distinguishable from other contributions
for relatively large rapidity separation between the $\phi$ mesons.
Hence, to study this type of mechanism one should investigate events with
rather large four-kaon invariant masses, outside of the region of resonances.
These events are then ``three-gap events'':
proton--gap--$\phi$--gap--$\phi$--gap--proton.
Experimentally, this should be a clear signature.
	\item
Clearly, an experimental study of CEP of a $\phi$-meson pair 
should be very valuable for clarifying the status of the odderon.
At least, it should be possible to derive an upper limit 
on the odderon contribution in this reaction.

\end{itemize}

\end{document}